\documentclass[aps,preprint,amsmath,amssymb,showpacs,amsfonts,prd,nofootinbib]{revtex4}
\usepackage{epsfig}
\usepackage{graphicx}
\usepackage{dcolumn}
\usepackage{bm}
\usepackage{amsmath, amsthm, amssymb}
\usepackage{color}

\newcommand{\bea}{\begin{eqnarray}}
\newcommand{\eea}{\end{eqnarray}}

\begin{document}
\title{Jeans Instability in Superfluids}
\author{Itamar Hason}
\author{Yaron Oz}
\affiliation{Raymond and Beverly Sackler School of
Physics and Astronomy, Tel-Aviv University, Tel-Aviv 69978, Israel}
\date{\today}
\begin{abstract}
We analyze the effect  of a gravitational field on the sound modes of superfluids.
We derive an instability condition that generalizes the well known Jeans instability of the sound mode in normal fluids. We discuss
potential experimental implications.

\end{abstract}


\maketitle

\section{Introduction}

The Jeans instability of the sound mode in normal fluids that is caused by a gravitational field is a well known phenomenon  \cite{jeans}.
It exhibits itself in astrophysical scenarios of aggregation of masses  and galaxy formation.
The Jeans dispersion relation of a normal fluid sound mode is obtained by linearizing the fluid equations in the presence of a gravitational field.
It modifies the sound mode dispersion relation
$ \omega^2 = u_1^2 \vec{k}^2$ 
to
\begin{equation}
 \omega^2 = u_1^2 \vec{k}^2 - 4\pi G \rho \ . 
 \label{Jeans}
\end{equation}
$\omega$ is the frequency, $\vec{k}$ is
the momentum vector, 
$u_1$ is the speed of sound in a normal fluid $u_1^2 = \frac{\partial p}{\partial \rho}$ evaluated at fixed entropy, $p$ is the pressure,
$\rho$ the local mass density and $G$ is the gravitational coupling.
Instability occurs when the RHS of (\ref{Jeans}) is negative
\begin{equation}
 \vec{k}^2  <  \frac{4\pi G \rho}{u_1^2} \ . 
 \label{Jeansins}
\end{equation}
It determines a Jeans length scale $\lambda_J$, where
all scales larger than that being unstable to a gravitational collapse.

In this letter we will analyze the effect of a gravitational field  on the sound modes in superfluids.
Superfluids are quantum fluids, i.e.  fluids at temperatures close to zero where quantum effects are of primary importance \cite{superfluid}. 
 They exhibit remarkable properties such as the ability to flow without viscosity in narrow capillaries.
 Superfluidity/superconductivity is also expected to be realized in Neutron stars matter  (see e.g. \cite{Lombardo:2000ec}), as well as in high density phases of
 QCD \cite{Alford:1997zt}.

 The hydrodynamic description of superfluids consists of two separate motions, a normal flow and a super flow with densities $\rho_n$ and $\rho_s$ respectively, 
 $\rho_n + \rho_s = \rho$ \cite{lan, Landau}.
The super flow moves without viscosity, 
the normal flow is viscous, and the two flows do not exchange momentum between them.
 We denote by $\vec{v}_n$ and $\vec{v}_s$ the velocities of the normal and super flows, respectively. 
 The superfluid velocity $\vec{v}_s$ corresponds to the gradient of the condensate phase that breaks spontaneously the global/local symmetry, which results in 
superfluidity/superconductivity, respectively.
Thus, the superfluid motion is irrotational
\begin{equation} \label{eq:irrotational_superfluid_flow}
 \vec{\nabla} \times \vec{v}_s = 0  \ .
\end{equation}

There are two sound modes in superfluids: the first sound $u_1$ which is a density wave as in normal fluids, and the second
 sound $u_2$, which is a temperature wave and is unique to superfluids. Its dispersion relation reads
 \begin{equation}
 \omega^2 =  u_2^2 \vec{k}^2, ~~~~~u_2^2 = \frac{\rho_s s^2}{\rho_n \frac{\partial s}{\partial T}}  \  ,
 \label{sound2}
\end{equation}
where $T$ is the temperature, $s$ is the entropy density 
per particle and the derivative is taken at fixed pressure. As expected, the second sound vanishes in the limit $\rho_s\rightarrow 0$.
 In general the waves can be a superposition of the two.
 
As we will show, in the absence of density fluctuations the pure second sound (\ref{sound2}) remains stable, while density fluctuations imply
a stability criterion for both sound modes and their superpositions.
It reads
\begin{equation}
\vec{k}^2 <  \frac{4\pi G \rho}{u_1^2+\frac{1}{\left( { 1-J } \right)}u_2^2} \ ,
 \label{jeansins3}
\end{equation}
where $J = \frac{\frac{\partial s}{\partial p} \frac{\partial \rho}{\partial T}}{\frac{\partial s}{\partial T} \frac{\partial \rho}{\partial p}}$.
This condition generalizes the ordinary Jeans instability
(\ref{Jeansins}) and reduces to it in the limit $\rho_s \rightarrow 0$. 
The RHS of (\ref{jeansins3}) depends on the 
details of the system thermodynamic properties. An interesting  limit to take is that
of very low temperatures $T\rightarrow 0$, where the super flow component is dominant.
In this  limit one has $u_1^2=3u_2^2$ \cite{Landau} and $J\rightarrow 0$ \cite{data}.
Thus, at very low temperaure we have a larger Jeans length than that of a normal
fluid by a factor of $\frac{2}{\sqrt{3}}$.

The letter is organized as follows. We will introduce a gravitational potential to the superfluid hydrodynamics equations and study
fluctuations at the linear order. We will then analyze the instability conditions and derive (\ref{jeansins3}), which is our main result.
Finally, we will discuss potential experimental implications.

\section{Gravitational Instability in Superfluid Hydrodynamics}

We will consider the evolution equations of a superfluid in the presence of a gravitational field. We denote the gravitational potential by $\phi$. It satisfies Gauss's law
\begin{equation}
\nabla^2\phi = 4\pi G \rho \ .
\end{equation}

\subsection{Self-gravitating superfluid hydrodynamics}

 We will consider an ideal superfluid.
The superfluid density current reads
\begin{equation}
 \vec{j} = \rho_n \vec{v}_n + \rho_s \vec{v}_s \ , 
\end{equation}
and it satisfies a continuity equation, which is not affected by the gravitational field
\begin{equation}
 \frac{\partial \rho}{\partial t} + \vec{\nabla} \cdot \vec{j} = 0 \ .
\end{equation}
Similarly, the entropy conservation is not affected by the gravitational field and reads
\begin{equation}
 \frac{\partial}{\partial t} \left( {\rho s} \right) + \left( {\rho s} \right) \vec{\nabla} \cdot \vec{v}_n = 0 \ .
 \end{equation}
 Note, that entropy is carried only by the normal flow and not by the super flow.
 
The energy and momentum conservation equations, however,  are modified in the presence of the gravitational field and take the form
\begin{equation}
\frac{\partial j_i}{\partial t} + \frac{\partial \Pi_{ik}}{\partial x_k} + \rho \frac{\partial\phi}{\partial x_i} = 0 \ ,
\end{equation}
\begin{equation}
\frac{\partial E}{\partial t} + \vec{\nabla} \cdot \vec{Q} +\vec{j} \cdot \vec{\nabla}\phi= 0 \ .
\end{equation}
$\Pi_{ik}$ is the momentum flux tensor and  $\vec{Q}$ is the energy flux as calculated without a gravitational field in \cite{Landau}.

In order to derive the the modified equations in the presence of the gravitational field, one uses thermodynamics  and the Galilean principle \cite{Landau}.
One denotes by $K_0$ the reference frame, where the super flow velocity is zero. The velocity of the normal flow
in this frame is $\vec{v}_n - \vec{v}_s$.
In this frame one has 
\begin{equation}
 E_0  = -p + T\rho s + \mu\rho + \phi\rho + \left( { \vec{v}_n - \vec{v}_s } \right) \cdot \vec{j}_0 \ ,
\end{equation}
and 
\begin{equation}
  d\left( { \mu + \phi } \right) = -s dT + \frac{1}{\rho} dp - \frac{\rho_n}{\rho} \left( { \vec{v}_n - \vec{v}_s } \right) d \left( { \vec{v}_n - \vec{v}_s } \right) \ .
\end{equation}
$E_0$  and $j_0$ are the energy and density current in the system $K_0$, respectively.
There are two  effects of the gravitational field. First,  to replace  the chemical potential $\mu$ by $\mu + \phi$. Second,
to introduce a new term in the fluid flow: in addition to the force term $-\vec{\nabla} p$, we have $-\rho\vec{\nabla}\phi$.

Finally, the super flow being a potential flow satisfies
\begin{equation}
 \frac{\partial \vec{v}_s}{\partial t} + \vec{\nabla} \left( {  \frac{1}{2}{\vec{v}_s}^2 + \mu  + \phi } \right) = 0 \ .
 \end{equation}
 
 It is straightforward to see that the hydrodynamics equations together with the thermodynamic relations constitute a complete
 set that determines all the charge densities and velocities.
 
\subsection{Self-gravitating superfluid sound} 
 
Linearizing the above ideal superfluid hydrodynamics equations in a presence of the gravitational field we get  
\begin{equation} \label{eq:linearized_superfluid_hydrodynamics_gravity}
 \begin{aligned}
  & \frac{\partial \rho}{\partial t} + \vec{\nabla} \cdot \vec{j} = 0 \\
  & \frac{\partial}{\partial t} \left(\rho s \right) + \left(\rho s \right) \vec{\nabla} \cdot \vec{v}_n = 0 \\
  & \frac{\partial \vec{j} }{\partial t} + \vec{\nabla} {p} + \rho \vec{\nabla} \phi= 0 \\
  & \frac{\partial \vec{v}_s}{\partial t} + \vec{\nabla} \mu + \vec{\nabla} \phi = 0  \  , \\
 \end{aligned}
\end{equation}
from which we derive
\begin{equation}
 \begin{aligned}
  & \frac{\partial^2\rho}{\partial t^2} = \nabla^2 p +\rho\nabla^2\phi , \\ \label{eq:superfluid_wave_first_gravity} 
  & \frac{\partial^2 s}{\partial t^2} = \frac{\rho_s}{\rho_n} s^2 \nabla^2 T + \frac{\rho_s}{\rho_n}s\nabla^2\phi  \ . \\
 \end{aligned}
\end{equation}

Perturbing\footnote{As in the ordinary Jeans instability analysis, we perturb around a stable state and therefore ignore the zeroth order contribution of the Laplacian of the gravitational potential $\nabla^2\phi_0$.} around \eqref{eq:superfluid_wave_first_gravity}, using Gauss's law for the gravitational potential, and expressing all quantities in terms of the pressure $p$ and the temperature $T$ we get
\begin{equation}
 \begin{aligned}
  & \frac{\partial \rho}{\partial p} \frac{\partial^2 \delta p}{\partial t^2} + \frac{\partial \rho}{\partial T} \frac{\partial^2 \delta T}{\partial t^2} = \nabla^2 {\delta p} + 4\pi G \rho \left( { \frac{\partial\rho}{\partial p}\delta p + \frac{\partial\rho}{\partial T}\delta T} \right)\\
  & \frac{\partial s}{\partial p} \frac{\partial^2 \delta p}{\partial t^2} + \frac{\partial s}{\partial T} \frac{\partial^2 \delta T}{\partial t^2} = \frac{\rho_s}{\rho_n}  s^2 \nabla^2 \delta T + 4\pi G \frac{\rho_s}{\rho_n} s \left( { \frac{\partial\rho}{\partial p}\delta p + \frac{\partial\rho}{\partial T}\delta T} \right) \ .
 \end{aligned}
\end{equation}

This system of equations for a wave of the form $\exp\left({i\vec{k}\cdot\vec{x}-i\omega t}\right)$ reads
\begin{equation} \label{eq:gravitational_superfluid_sound}
 \begin{aligned}
  & \left( {\frac{\partial \rho}{\partial p} \left( {\omega^2 + 4\pi G \rho} \right) - \vec{k}^2} \right) \delta p + \frac{\partial \rho}{\partial T} \left( { \omega^2 + 4\pi G \rho } \right) \delta T = 0 \\
  & \left( { \frac{\partial s}{\partial p}\omega^2 +\frac{\partial \rho}{\partial p} \cdot 4\pi G \frac{\rho_s}{\rho_n}s}  \right) \delta p + \left( { \frac{\partial s}{\partial T} \omega^2 - \frac{\rho_s}{\rho_n}s^2 \vec{k}^2 + \frac{\partial \rho}{\partial T} \cdot 4\pi G \frac{\rho_s}{\rho_n}s} \right) \delta T = 0  \ . \\
 \end{aligned}
\end{equation}
In the absence of a gravitational field ($G\rightarrow 0$) the equations reduce to the known analysis of superfluid sound modes \cite{Landau}.

In the general case, for a solution of these system of equations we require the determinant to vanish
\begin{equation}
H \omega^4 + 
 \left(4\pi G H \rho - \left(\frac{\partial \rho}{\partial p} \frac{\rho_s}{\rho_n} s^2 + \frac{\partial s}{\partial T}  \right) 
 \vec{k}^2 \right) \omega^2  
 + \left( \frac{\rho_s}{\rho_n}s^2 \vec{k}^2 - 4\pi G \frac{\rho_s}{\rho_n} s \left(\frac{\partial \rho}{\partial p}\rho s + \frac{\partial \rho}{\partial T} \right) \right) 
 \vec{k}^2  = 0 \ ,
\end{equation}
where $H = \frac{\partial \left( \rho, s \right)}{\partial \left( p, T  \right) }$ is the Jacobian for changing variables from $(\rho,s)$ to
$(p,T)$, and we assume that $H > 0$.
Solving for $\omega^2$ we get
\begin{eqnarray}
 2 H \omega^2 &=&  - \left(4\pi G H \rho - \left(\frac{\partial \rho}{\partial p} \frac{\rho_s}{\rho_n}s^2 + \frac{\partial s}{\partial T} \right)\vec{k}^2  \right)
  \nonumber \\
 &\pm&  \sqrt{\left(4 \pi G  H \rho + \left(\frac{\partial \rho}{\partial p}\frac{\rho_s}{\rho_n}s^2 - \frac{\partial s}{\partial T} \right)\vec{k}^2  \right)^2 + 
 4 \frac{\partial \rho}{\partial T} \frac{\rho_s}{\rho_n}  s\vec{k}^2 \left(4 \pi G H + \frac{\partial s}{\partial p} s \vec{k}^2 \right) } \ . \label{sol}
\end{eqnarray}
In the absence of a gravitational field, we have the plus and minus sign solutions reducing  to the first and second sounds, respectively.

Consider the leading effect of the super flow in the dimensionless parameter $\frac{\rho_s}{\rho_n}\ll 1$.  At first order we get
\begin{eqnarray}
 2 H \omega^2 =&  - \left(4\pi G H \rho - \frac{\partial s}{\partial T} \vec{k}^2  \right) \pm  \left(4 \pi G  H \rho - \frac{\partial s}{\partial T}\vec{k}^2  \right)
 \nonumber \\
 & + \frac{\rho_s}{\rho_n}\left( { \frac{\partial \rho}{\partial p} s^2 \pm \frac{\partial \rho}{\partial p} s^2 \pm \frac{2\frac{\partial \rho}{\partial T}  s\vec{k}^2 \left(4 \pi G H + \frac{\partial s}{\partial p} s \vec{k}^2 \right)}{\left(4 \pi G  H \rho - \frac{\partial s}{\partial T}\vec{k}^2  \right)}} \right) \ .
\end{eqnarray}
Therefore, we have the first sound solution
\begin{eqnarray}
 H \omega^2 = - \left(4\pi G H \rho - \frac{\partial s}{\partial T} \vec{k}^2  \right) - \frac{\rho_s}{\rho_n}\left( { \frac{\frac{\partial \rho}{\partial T}  s\vec{k}^2 \left(4 \pi G H + \frac{\partial s}{\partial p} s \vec{k}^2 \right)}{\left(4 \pi G  H \rho - \frac{\partial s}{\partial T}\vec{k}^2  \right)}} \right) \ ,
\end{eqnarray}
and the second sound solution
\begin{eqnarray}
 H \omega^2 = \frac{\rho_s}{\rho_n}\left( { \frac{\partial \rho}{\partial p} s^2 + \frac{\frac{\partial \rho}{\partial T}  s\vec{k}^2 \left(4 \pi G H + \frac{\partial s}{\partial p} s \vec{k}^2 \right)}{\left(4 \pi G  H \rho - \frac{\partial s}{\partial T}\vec{k}^2  \right)}} \right)  \ .
\end{eqnarray}

\subsection{Comparison to the Non-Gravitational Case}

It is instructive to compare special solutions to the superfluid sound equations in the presence of the gravitational interaction (\ref{eq:gravitational_superfluid_sound})
to the non-gravitating case. 

\subsubsection{The Pure Second Sound Limit}

In the limit  $\frac{\partial \rho}{\partial T} = 0$, the non-gravitational sound equations (setting $G=0$ in (\ref{eq:gravitational_superfluid_sound})) read
\begin{equation} \label{eq:simplest_non_gravitational_superfluid_sound}
 \begin{aligned}
  & \left( {\frac{\partial \rho}{\partial p} \omega^2 - \vec{k}^2} \right) \delta p = 0 \\
  & \left( { \frac{\partial s}{\partial p}\omega^2 }  \right) \delta p + \left( { \frac{\partial s}{\partial T} \omega^2 - \frac{\rho_s}{\rho_n}s^2 \vec{k}^2 } \right) \delta T = 0 \ .
 \end{aligned}
\end{equation}
One has a solution where $\delta p = 0$, $\delta T$ arbitrary, and the dispersion relation (\ref{sound2}) of the pure second sound describing the temperature wave.
There is another solution, where the dispersion relation is that of a first sound mode,
and the pressure and temperature variations are related by
\begin{equation}
 \left( { \frac{\partial s}{\partial p} }  \right) \delta p = \left( { \frac{\rho_s}{\rho_n}s^2 \frac{\partial \rho}{\partial p} - \frac{\partial s}{\partial T} } \right) \delta T \ .
\label{one}
\end{equation}

When we introduce gravity, still having $\frac{\partial \rho}{\partial T} = 0$, we get
\begin{equation} \label{eq:simplest_gravitational_superfluid_sound}
 \begin{aligned}
  & \left( {\frac{\partial \rho}{\partial p} \left( {\omega^2 + 4\pi G \rho} \right) - \vec{k}^2} \right) \delta p = 0 \\
  & \left( { \frac{\partial s}{\partial p}\omega^2 +\frac{\partial \rho}{\partial p} \cdot 4\pi G \frac{\rho_s}{\rho_n}s}  \right) \delta p + \left( { \frac{\partial s}{\partial T} \omega^2 - \frac{\rho_s}{\rho_n}s^2 \vec{k}^2} \right) \delta T = 0 \ .
 \end{aligned}
\end{equation}
We see that the pure second sound solution (\ref{sound2}) is not affected by gravity.
This is expected since the pure second sound is characterized by no variation of the density, thus gravity does not affect it.

However, the first sound solution is changed. The dispersion relation is the same as the Jeans dispersion relation (\ref{Jeans}),
but the pressure and temperature variations are related by a different formula
\begin{equation} 
 \left( { \frac{\partial s}{\partial p}\omega^2 +\frac{\partial \rho}{\partial p} \cdot 4\pi G \frac{\rho_s}{\rho_n}s}  \right) \delta p = \left( { \left( { \frac{\partial \rho}{\partial p} \frac{\rho_s}{\rho_n}s^2 - \frac{\partial s}{\partial T} } \right) \omega^2 + \frac{\partial \rho}{\partial p} \cdot  4\pi G  \frac{\rho_s}{\rho_n} \rho s^2 } \right) \delta T \ .
\label{two}
\end{equation}
Comparing (\ref{one}) and (\ref{two}), we  see a frequency dependence of the linear relation between pressure and temperature (\ref{two}), which vanishes as
$\rho_s\rightarrow 0$.

\subsubsection{The Pure First Sound Limit}
Consider next the classical pure first sound limit, $\frac{\partial s}{\partial p} = 0$. The non-gravitational sound equations in this case are (setting $G=0$ and $\frac{\partial s}{\partial p} = 0$ in (\ref{eq:gravitational_superfluid_sound})) :
\begin{equation} \label{eq:second_simplest_non_gravitational_superfluid_sound}
 \begin{aligned}
  & \left( {\frac{\partial \rho}{\partial p} \omega^2 - \vec{k}^2} \right) \delta p + \left( { \frac{\partial \rho}{\partial T} \omega^2   } \right) \delta T = 0 \\
  & \left( { \frac{\partial s}{\partial T} \omega^2 - \frac{\rho_s}{\rho_n}s^2 \vec{k}^2 } \right) \delta T = 0 \ .
 \end{aligned}
\end{equation}
We have a solution where $\delta T = 0$, $\delta p$ arbitrary, and the dispersion relation 
of the pure first sound.
There is another solution, where the dispersion relation is (\ref{sound2}), 
and the pressure and temperature variations are related by
\begin{equation}
 \left( { \frac{\partial s}{\partial T} \frac{\rho_n}{\rho_s} \frac{1}{s^2} - \frac{\partial \rho}{\partial p} } \right) \delta p = 
 \left( { \frac{\partial \rho}{\partial T} } \right) \delta T  \ .
\end{equation}

When we introduce gravity, still with $\frac{\partial s}{\partial p} = 0$, we get
\begin{equation} \label{eq:second_simplest_gravitational_superfluid_sound}
 \begin{aligned}
  & \left( {\frac{\partial \rho}{\partial p} \left( {\omega^2 + 4\pi G \rho} \right) - \vec{k}^2} \right) \delta p + \frac{\partial \rho}{\partial T} \left( { \omega^2 + 4\pi G \rho } \right) \delta T = 0 \\
  & \left( { \frac{\partial \rho}{\partial p} \cdot 4\pi G \frac{\rho_s}{\rho_n}s}  \right) \delta p + \left( { \frac{\partial s}{\partial T} \omega^2 - \frac{\rho_s}{\rho_n}s^2 \vec{k}^2 + \frac{\partial \rho}{\partial T} \cdot 4\pi G \frac{\rho_s}{\rho_n}s} \right) \delta T = 0 \ .
 \end{aligned}
\end{equation}
We see that the pure first sound solution is not pure anymore.
This is in contrast to the pure second sound solution that was not modified by gravity.

\subsection{General Instability Conditions}
In the following we will analyze the instability conditions arising from the roots (\ref{sol}).

\subsubsection{Identical roots}

Consider the case where the two solutions (\ref{sol}) are identical.
In this case we have the same dispersion relation for the first and second sound and we obtain
the instability condition
\begin{equation}
\vec{k}^2 <  \frac{4\pi G H \rho}{\left( { \frac{\partial \rho}{\partial p} \frac{\rho_s}{\rho_n}s^2 + \frac{\partial s}{\partial T} } \right)}  \ .
 \label{jeansins2}
\end{equation}
Dividing by $H$, we can recast (\ref{jeansins2}) as (\ref{jeansins3}). 
The constraint (\ref{jeansins3}) is a generalization of 
(\ref{Jeansins}) to superflows, and is valid for both the first and second sounds.
It reduces to  (\ref{Jeansins})  in the limit $\rho_s \rightarrow 0$. 
In limit  $T\rightarrow 0$, $u_1^2=3u_2^2$ \cite{Landau}, and one can see from the data presented in \cite{data} that $J\rightarrow 0$. Thus we have that the Jeans length of superfluid is larger by a factor of
$\frac{2}{\sqrt{3}}$ than that of a normal fluid.

\subsubsection{Different roots}

The conditions for instability of the two sound modes 
are the upper bound on the momentum (\ref{jeansins3}) as well as a lower bound
\begin{equation} \label{eq:double_instability_4}
\vec{k}^2  > 4 \pi G \rho \left( \frac{\partial \rho}{\partial p} + \frac{\partial \rho}{\partial T}  \frac{1}{\rho s} \right)  \  ,
\end{equation}
which together  imply a constraint on the system independent of the gravitational field
\begin{equation} 
\left( \frac{\partial \rho}{\partial p} + \frac{\partial \rho}{\partial T}  \frac{1}{\rho s} \right) \left( { \frac{\partial \rho}{\partial p} \frac{\rho_s}{\rho_n}s^2 + \frac{\partial s}{\partial T} } \right) < H \ .  \label{H}
\end{equation}


\subsection{Discussion}

As the ordinary Jeans instability has been observed experimentally, it is natural to inquire how can we observe the generalized instability conditions for superfluids.
It is probably unlikely that current condensed matter experiments can detect the gravitational instability of superfluids. 
A potential experimental laboratory could be Neutron stars.
Superfluidity is expected to occur both in the inner crust and the core of Neutron stars as a consequence of
the formation of Cooper pairs of Neutrons that break spontaneously the $U(1)$ Baryon symmetry \cite{Lombardo:2000ec}.
Superfluidity has been invoked in order to explain, for instance, pulsar glitches  \cite{anderson}.
Therefore the gravitational instability seems relevant to the dynamics of Neutron stars, such as its rotational 
and vibrational oscillations.
It would be interesting to work out the details of the gravitational instability effects in this framework.

\section*{Acknowledgements}
This work is supported in part by the I-CORE program of Planning and Budgeting Committee (grant number 1937/12), and by the US-Israel Binational Science Foundation.


\begin{thebibliography}{References}

\bibitem{jeans}
J. H. Jeans, "The Stability of a Spherical Nebula". Philosophical Transactions of the Royal Society A {\bf 199}, 1 (1902).

\bibitem{superfluid}
P. Kapitza,  "Viscosity of Liquid Helium Below the $\lambda$-Point", Nature {\bf 141},  74 (1938).


\bibitem{Lombardo:2000ec} 
  U.~Lombardo and H.~J.~Schulze,
  Lect.\ Notes Phys.\  {\bf 578}, 30 (2001)
  [astro-ph/0012209].

\bibitem{Alford:1997zt} 
  M.~G.~Alford, K.~Rajagopal and F.~Wilczek,
  ``QCD at finite baryon density: Nucleon droplets and color superconductivity,''
  Phys.\ Lett.\ B {\bf 422}, 247 (1998)
  [hep-ph/9711395].

\bibitem{lan}
L. D. Landau, "The Theory of Superfluidity of Helium II",  J. Phys. USSR, {\bf 5},  71(1941).

\bibitem{Landau}
  E.~M.~Lifshitz and L.~D.~Landau,
  ``Course of Theoretical Physics, volume 6, Fluid Mechanics'',
  Pergamon Press (1959).

\bibitem{data} J. Brooks and R. Donnelly, "The Calculated Thermodynamic Properties of Superfluid Helium-4", J. Phys. Chem. Data, {\bf 6}, 1 (1977).

\bibitem{anderson}
P.W. Anderson and N. Itoh, "Pulsar glitches and restlessness as a hard superfluidity phenomenon",
Nature {\bf 256}, 25 (1975);.

\end{thebibliography}
\end{document}